%Paper: cond-mat/9406109
%From: Ely Ben-Naim <ebn@buphyk.bu.edu>
%Date: Mon, 27 Jun 1994 17:59:43 -0400

%----------------------------TeX manuscript---------------------------
\def\cp #1 #2 #3 {{\sl Chem.\ Phys.} {\bf #1}, #2 (#3)}
\def\jetp #1 #2#3 {{\sl Sov.\ Phys.\ JETP} {\bf #1}, #2 (#3)}
\def\jpa #1 #2 #3 {{\sl J. Phys.\ A} {\bf #1}, #2 (#3)}
\def\jcp #1 #2 #3 {{\sl J.\ Chem.\ Phys.} {\bf #1}, #2 (#3)}
\def\jpc #1 #2 #3 {{\sl J.\ Phys.\ Chem.} {\bf#1}, #2 (#3)}
\def\jsp #1 #2 #3 {{\sl J.\ Stat.\ Phys.} {\bf #1}, #2 (#3)}
\def\jdep #1 #2 #3 {{\sl J.\ de Physique I} {\bf #1}, #2 (#3)}
\def\pra #1 #2 #3 {{\sl Phys.\ Rev.\ A} {\bf #1}, #2 (#3)}
\def\prb #1 #2 #3 {{\sl Phys.\ Rev.\ B} {\bf #1}, #2 (#3)}
\def\pre #1 #2 #3 {{\sl Phys.\ Rev.\ E} {\bf #1}, #2 (#3)}
\def\prl #1 #2 #3 {{\sl Phys.\ Rev.\ Lett.} {\bf #1}, #2 (#3)}
\def\rmp #1 #2 #3 {{\sl Rev.\ Mod.\ Phys.} {\bf #1}, #2 (#3)}
\def\zpc #1 #2 #3 {{\sl Z. Phys.\ Chem.} {\bf #1}, #2 (#3)}
\def\zw #1 #2 #3 {{\sl Z. Wahrsch.\ verw.\ Gebiete} {\bf #1}, #2 (#3)}

%%%%%%%%%%%%%%%%%%%%%%%%%%%%%%%%%%%%%%%%%%%%%%%%%%%%%%%%%%%%%%%%%%%%%%%%%%%%%%

\def\gtwid{\mathrel{\raise.3ex\hbox{$>$\kern-.75em\lower1ex\hbox{$\sim$}}}}
\def\ltwid{\mathrel{\raise.3ex\hbox{$<$\kern-.75em\lower1ex\hbox{$\sim$}}}}
\def\eg{{\it e.\ g.}}\def\ie{{\it i.\ e.}}

\def\pd#1#2{{\partial #1\over\partial #2}}      %partial derivative
\def\p2d#1#2{{\partial^2 #1\over\partial #2^2}} %second partial derivative
\def\td#1#2{{d #1\over d #2}}      %total derivative
\def\t2d#1#2{{d^2 #1\over d #2^2}} %second total derivative
\def\ith{{$i^{\rm th}$}}

\def\kth{{$k^{\rm th}$}}

\newcount\refnum\refnum=0  %automatic, non-labelled references
\def\refi{\smallskip\global\advance\refnum by 1\item{\the\refnum.}}
\newcount\rfignum\rfignum=0  %automatic, non-labelled figure labels in captions
\def\rfigi{\medskip\global\advance\rfignum by 1\item{Figure \the\rfignum.}}

\newcount\eqnum \eqnum=0  %automatic, non-labelled equation numbering
\def\eqnoi{\global\advance\eqnum by 1\eqno(\the\eqnum)}
\def\eqnai{\global\advance\eqnum by 1\eqno(\the\eqnum {\rm a})}
\def\eqnbi{\eqno(\the\eqnum {\rm b})}
\def\back#1{{\advance\eqnum by-#1 Eq.~(\the\eqnum)}}%refer to one equation back
\def\last{Eq.~(\the\eqnum)}                   %refer to the current equation

%%%%%%%%%%%%%%%%%%%%%%%%%%%%%%%%%%%%%%%%%%%%%%%%%%%%%%%%%%%%%%%%%%%%%%%%%%%%%%

\def\pvt{P(D,t)}
\def\pvT{P(D,T)}
\def\pv0{P(D,0)}
\def\pvvt{P(D',t)}
\def\pvvT{P(D',T)}

\def\ct{c(t)}
\def\cT{c(T)}
\def\ctt{c(t')}
\def\cTT{c(T')}
\def\cta{{c\sim t^{-\a}}}
\def\dtb{{\dav \sim t^{-\b}}}
\def\dav{\langle D\rangle}
\def\dm{D_{\rm min}}
\def\a{\alpha}
\def\b{\beta}
\def\m{\mu}
\def\d{\delta}
\def\e{\epsilon}
\def\g{\gamma}
\def\G{\Gamma}

\def\CP{\hbox{{$\cal P$}}}

%%%%%%%%%%%%%%%%%%%%%%%%%%%%%%%%%%%%%%%%%%%%%%%%%%%%%%%%%%%%%%%%%%%%%%%%%%%%%%

\def\indd{{\int \limits_0^{\infty}dD'}}
\def\ind{{\int \limits_0^{\infty}dD}}
\def\intt{{\int \limits_0^t dt'}}
\def\inTT{{\int \limits_0^T dT'}}

\overfullrule=0pt
\magnification 1200
\baselineskip=22 true bp
%\baselineskip=26 true bp

%%%%%%%%%%%%%%%%%%%%%%%%%%%%%%%%%%%%%%%%%%%%%%%%%%%%%%%%%%%%%%%%%%%%%%%%%%%%%%%

\centerline{\bf Kinetics of Heterogeneous Single-Species Annihilation}
\bigskip
\centerline{\bf P.~L.~Krapivsky, E.~Ben-Naim, and S.~Redner}
\smallskip
\centerline{Center for Polymer Studies and Department of Physics}
\centerline{Boston University, Boston, MA 02215}
\vskip 1in

\centerline{ABSTRACT}
{\smallskip\narrower\noindent We investigate the kinetics of
diffusion-controlled {\it heterogeneous} single-species annihilation,
where the diffusivity of each particle may be different.  The
concentration of the species with the smallest diffusion coefficient
has the same time dependence as in {\it homogeneous} single-species
annihilation, $A+A \to 0$.  However, the concentrations of more mobile
species decay as power laws in time, but with non-universal exponents
that depend on the ratios of the corresponding diffusivities to that
of the least mobile species.  We determine these exponents both in a
mean-field approximation, which should be valid for spatial dimension
$d>2$, and in a phenomenological Smoluchowski theory which is
applicable in $d<2$.  Our theoretical predictions compare well with
both Monte Carlo simulations and with time series expansions.
}
{

\bigskip\narrower\noindent P. A. C. S. Numbers: 02.50.-r, 05.40.+j, 82.20.-w,
82.20.Wt
}

\vfill\eject

%%%%%%%%%%%%%%%%%%%%%%%%%%%%%%%%%%%%%%%%%%%%%%%%%%%%%%%%%%%%%%%%%%%%%%%%%%

\medskip\centerline{\bf I. Introduction}\medskip

The kinetics of diffusion-controlled single-species annihilation, $A+A
\to 0$, when each particle has the same diffusion coefficient is now well
understood [1].  For spatial dimension $d>2$, the kinetics may be
accounted for by the rate equation which predicts that the density
decays as $t^{-1}$ in the long-time limit.  For $d\leq2$, various
phenomenological approaches predict that the density decays as
$t^{-d/2}$, but with logarithmic corrections appearing in $d=2$.
Accompanying this relatively slow kinetics is a spatial organization
in which the probability of finding particles at small separations is
reduced compared to a random distribution.  In one dimension, exact
solutions, either based on an occupation number formalism [2], or by
mapping the reaction onto the kinetic Ising-Glauber model [3], provide
definitive results about this spatial organization and the reaction
kinetics.

Our goal in this paper is to describe the kinetics of {\it heterogeneous}
single-species annihilation, which is defined by the reaction scheme
$A_i+A_k\buildrel K_{i,j}\over\longrightarrow 0$.  Here $A_i$ denotes the
\ith\ species and the reaction rate matrix $K_{i,j}$ is a function of the
diffusivities of the two reacting species.  Although we refer to
different species in the context of their diffusivity, the reaction
itself is single-species annihilation with distinct rates for
different reaction channels.  We will consider reactant diffusivities
which are drawn from a probability distribution.  Such a situation
arises naturally when the reactants have different masses.  As we
shall show, the kinetics of the heterogeneous system is considerably
richer than that of the homogeneous analog.  Similar behavior was also
encountered in heterogeneous single-species annihilation with
ballistic particle motion [4], where the kinetics depends in an
essential way on the form of the initial distribution of velocities.
The present investigation is a natural counterpart of this earlier
work for diffusive single-species annihilation.

When the number of species is finite (\ie, the density distribution,
$P(D,t)$, contains a finite number of discrete peaks), the rate
equations predicts that the least mobile species (with diffusivity
$\dm$) decays as $t^{-1}$, as in homogeneous annihilation.  However,
the more mobile species each decay at a faster power-law rate with an
associated exponent that depends on the diffusivity ratio between the
more mobile and the slowest species.  When $P(D,t)$ is continuous but
with $\dm>0$, the rate equations again show that the least mobile
species predominates in the long time limit, and that the decay of the
more mobile species is described by non-universal power-law behavior.
However, the detailed form of $P(D,t=0)$ near $\dm$ contributes to
a logarithmic prefactor in the decay law.  If $\dm=0$, the kinetics
strongly depends on the initial conditions.  For an initial
distribution of diffusivities with a power law tail, $\pv0\sim
D^{\m}$, as $D \to 0$, the concentration and the average diffusion
coefficient decay as $\cta$ and $\dtb$, respectively, with $\a =
(2+2\m)/(3+2\m)$ and $\b = 1-\alpha=1/(3+2\m)$.  All of these
predictions are expected to apply for $d>2$.

For $d\leq 2$, we apply the Smoluchowski theory to account for the
reaction kinetics.  We first test this phenomenological approach on the
``impurity'' problem, in which there is a small concentration of
particles with one diffusion coefficient in a homogeneous background of
particles with a different diffusivity.  The Smoluchowski theory
predicts that the impurity species decays as a non-universal power-law
in time.  These predictions are in good agreement with numerical results
that arise from time-series expansions and Monte Carlo simulations in
one dimension.  Connections are also made with the exactly-soluble
problem of the decay of impurities in the related single-species
coalescence process.  We then generalize the Smoluchowski approach to
both continuous distributions of diffusivities and general spatial
dimension $d<2$.

The rest of this paper is organized as follows. In Section 2, we
derive basic results from the rate equations for both discrete and
continuous distributions of diffusivities.  In Section 3, we
investigate the reaction kinetics for $d\leq2$ by the Smoluchowski
approach.  This leads to phenomenological rate equations with
time-dependent reaction rates.  The predictions of this approach are
then outlined.  In Section 4, we present time series expansion and
Monte Carlo simulation results in one dimension to support our
theoretical findings.  Finally, we give a brief discussion in Section
5.

%%%%%%%%%%%%%%%%%%%%%%%%%%%%%%%%%%%%%%%%%%%%%%%%%%%%%%%%%%%%%%%%%%%%%%%%%%%

\bigskip\centerline{\bf II. Rate Equations Approach}\medskip

Consider first the mean-field rate equations for the case of two
distinct species, $A_1$ and $A_2$, with respective diffusivities $D_1$
and $D_2$.  The two species interact according to the bimolecular
processes $A_1+A_1 \to 0$, $A_1+A_2 \to 0$, and $A_2+A_2 \to 0$, with
respective rates $K_{11}$, $K_{12}=K_{21}$, and $K_{22}$.  The
corresponding rate equations are
$$
\td {c_1} t = -K_{11}c_1^2 -K_{12}c_1c_2,
\eqnai
$$
$$
\td {c_2} t = -K_{22}c_2^2 -K_{12}c_1c_2,
\eqno(1b)
$$
where $c_i$ denotes the concentration of the \ith\ species.

Smoluchowski suggested a simple way to relate the reaction rate
$K_{ij}$ to the diffusion coefficients $D_i$ and $D_j$, and radii
$R_i$ and $R_j$ of the reactants [5].  This derivation is based on
considering the low-density limit.  In the rest frame of a particle of
species $i$, it may be considered as a spherical stationary trap of
radius $R$ which is surrounded by cloud of $j$ particles which are
captured upon contact with the trap.  The reaction rate is identified
as the flux of particles of type $j$ to the trap under the boundary
conditions of absorption at the surface of the \ith\ particle and a
fixed concentration as $r\to\infty$.  Upon solving the diffusion
equation in three dimensions under these conditions, one
straightforwardly finds that this flux equals $K=4\pi DR$ in the
long-time limit (see, $\eg$, [6]).  If both species perform
independent Brownian motions with respective diffusion coefficients
$D_i$ and $D_j$, then the appropriate generalization of the reaction
rate is $K_{ij}= 4\pi(D_i+D_j)(R_i+R_j)$.  Assuming, for simplicity,
that $R_1=R_2=R$ and absorbing the numerical factor $8\pi R$ into the
overall time scale, we may re-write \last\ as,

$$
\td {c_1} t = -2D_1c_1^2 -(D_1+D_2)c_1c_2,
\eqnai
$$
$$
\td {c_2} t = -2D_2c_2^2 -(D_1+D_2)c_1c_2.
\eqno(2b)
$$

To solve these
equations, consider first the time dependence of the concentration ratio
$\psi=c_2/c_1$.  By defining the auxiliary variable $dy=c_1dt$, the rate
equation for the concentration ratio is
$$
\td \psi y = -(D_2-D_1) (\psi+\psi^2).
\eqnoi
$$
Because $\psi$ is a strictly decreasing function of $y$ for $D_2>D_1$, the
asymptotic behavior may be estimated by ignoring the quadratic term in
\last.  The corresponding solution is simply
$$
\psi(y) \sim e^{-(D_2-D_1)\int_0^tdt'\,c_1(t')}.
\eqnoi
$$
Two possibilities for the behavior of
$y=\int dt'\,c_1(t')$  as $t\to\infty$ can occur.
If $y$ reaches a finite limit as $t\to\infty$, then $c_2(t)\propto c_1(t)$;
further, $c_1(t)$ and $c_2(t)$ must decay as $t^{-1}$.
It is immediate to show that these two conditions are
incompatible with the initial rate equations.  On the other hand,
if $y\to\infty$ as $t\to\infty$, then $c_2/c_1\to 0$, asymptotically.
Thus in Eq.~(2a), $c_1c_2$ may be neglected compared to $c_1^2$
and the resulting approximation gives $c_1(t)\simeq (2D_1t)^{-1}$.
Using this in the equation for $\psi$, one ultimately finds
$$
c_2 \simeq C (2D_1t)^{-(1+\d)}, \quad C = {c_2(0)
\over c_1(0)}
\biggl[{1+c_2(0)/c_1(0) \over c_1(0)}\biggr]^\d
\eqnoi
$$
for $t \to \infty$, with $\d=(D_2-D_1)/2D_1$.
While the less mobile species decays as $t^{-1}$ in
the long-time limit, as in homogeneous single-species annihilation,
the more mobile species decays non-universally as $t^{-\a}$ with
$\a = 1+\d =(D_1+D_2)/2D_1$.

It is worth noting that when $\d$ is small, the asymptotic behavior
given in \last\  is reached only at very
long times.  To demonstrate this, consider, $\eg$,
equal initial concentrations of the two species.
Then solving Eq.~(3) with $\psi(0)=1$ and $0<\d \ll 1$
gives $\psi(y) = c_2/c_1 \simeq 1+\d\ln[c_1(y)/c_1(0)]$.
Substituting this into Eqs.~(2), one finds

$$
c_{1,2}(t) \simeq {1\over 4D_1t}\Bigl[1 \pm {\d \over 2}\ln(D_1t)\Bigr],
\eqnoi
$$
for $1 \ll t \ll e^{2/\d}$.  Consequently, the crossover time between
the intermediate asymptotics, \last, and the final asymptotics, \back1,
diverges as $e^{2/\d}$ for $\d \to 0$.

For a finite number of species, the kinetics is similar to
that of the two-species case.  Namely the concentration of the least
mobile species decays as $t^{-1}$, while the
more mobile species exhibit non-universal power-law decays with
associated exponents $(D_1+D_k)/2D_1$, where $D_k$ is the diffusion
coefficient of the \kth\ species.

When the diffusivities of the species are drawn from a continuous
distribution, the rate equation for $\pvt$, the concentration of
species with diffusivity $D$, becomes the integro-differential equation
$$
\eqalign{
\pd \pvt t &= -\pvt\indd\, (D+D')\,\pvvt \cr
&= -\pvt \bigl[D\,\CP_0(t)+\CP_1(t)\bigr].\cr}
\eqnoi
$$
Here $\CP_k(t)$ denotes the $k^{\rm th}$ moment of the diffusivity distribution
$\pvt$,
$$
\CP_k(t) = \ind \,D^k \,\pvt.
\eqnoi
$$
Note that the zeroth moment of the diffusivity distribution is just
the particle density, $\ct = \CP_0(t)$, while the average diffusion
coefficient is expressed in terms of the zeroth and first moment
by $\dav = \CP_1(t)/\CP_0(t)$.

Equations similar to \back1\ describe the kinetics of several
irreversible processes, such as diffusion-reaction aggregation [7],
ballistic annihilation [4], and ballistic aggregation [8].
For the aggregation problem, a reaction rate of the form $(D+D')$
is known as the ``sum'' kernel, for which
the rate equations are exactly soluble [7].  In heterogeneous
annihilation, the rate equations are also soluble by elementary
analysis.  A formal, but implicit solution to \back1\ is
$$
\pvt = \pv0 \exp\Biggl[-D\intt \,\CP_0(t') -\intt \,\CP_1(t')\Biggr].
\eqnoi
$$
Further, by
integrating \back2\ over $D$, the following differential equation
which relates the moments $\CP_0(t)$ and $\CP_1(t)$ is obtained
$$
\td {\CP_0} t = -2\CP_0\CP_1.
\eqnoi
$$
With the initial condition $\CP_0(t=0)=1$, which fixes the
scale of the initial distribution, the solution to \last\ is
$$
\CP_0(t) = \exp\Biggl[ - 2\intt \,\CP_1(t')\Biggr].
\eqnoi
$$
Combining Eqs.~(9) and (11) leads to a simplification in which
only the unknown zeroth moment, or particle
density $\ct = \CP_0(t)$, appears in the formal solution
$$
\pvt = \pv0 \,\sqrt{\ct}\, \exp\Biggl[-D\intt \,\ctt\Biggr].
\eqnoi
$$

A scaling analysis of this solution indicates that two types of
behavior can occur: one for initial diffusivity distributions with a
finite non-zero lower cutoff, $\dm>0$, and the other for situations
where $\dm = 0$.  For both cases, we consider only those initial
distributions which are homogeneous near $\dm$, namely, $\pv0\sim
(D-\dm)^\m$ as $D \to \dm$ with $\m > -1$ for normalizability.  This
restriction leads to mathematical tractability as well as being a
natural illustrative choice.

For $\dm = 0$, the average diffusion coefficient clearly decay to zero,
and under mild restrictions, the particle concentration also decays to
zero.  For an initial distribution of the form $\pv0 \sim D^\m$ as $D
\to 0$, it is natural to assume power-law decays in the average
concentration and average diffusivity: $\cta$ and $\dtb$ for $t \to
\infty$.  Correspondingly, the time-dependent diffusivity distribution
is expected to approach the scaling form,
$$
\pvt \simeq t^{\b-\a} \Phi(Dt^\b).
\eqnoi
$$
Upon substituting this scaling ansatz into \back1, one finds two
relations which then determine the exponents $\a$ and $\b$.  First, for
the argument of the exponential in \back1\ to be
dimensionless, if follows that $\a+\b=1$.  Similarly, for the prefactor to
be a function only of the scaling variable $x=Dt^\b$, the relation
$\a=2\b(1+\m)$ must hold.  These two conditions determine the dependence
of the fundamental exponents on $\mu$ ,
$$
\a = {2+2\m \over 3+2\m}, \quad \b = {1 \over 3+2\m},
\eqnoi
$$
while the scaling  function is
$$
\Phi(x) = x^\m e^{-x}.
\eqnoi
$$

Let us now justify the existence of the scaling ansatz by constructing
explicit solutions to \back3.  Consider the initial distribution
$\pv0=D^\m e^{-D}/\G(\m+1)$, where the numerical
factor normalizes the initial density to unity, for convenience.
Substituting this
initial distribution into \back3\ and then integrating the resulting
equation over $D$, one finds the simple differential equation
$$
\td u t = u^{-2\m-2},
\eqnoi
$$
for the auxiliary function $u(t)=1+\intt \ctt$.  With the initial condition,
$u(t=0)=1$, \last\ is readily solved, from which the
concentration is
$$
\ct = \td u t= \Bigl[1+(2\m+3)t\Bigr]^{-{2+2\m \over 3+2\m}}.
\eqnoi
$$
Substituting this in \back5, we  obtain for the diffusivity distribution
$$
\pvt={D^\m \over \G(\m+1)} \Bigl[1+(2\m+3)t\Bigr]^{-{1+\m \over 3+2\m}}
\exp\Biggl(-D \Bigl[1+(2\m+3)t\Bigr]^{1 \over 3+2\m}\Biggr).
\eqnoi
$$
As expected, this solution approaches the scaling form of \back5\
asymptotically, with the exponents and scaling function given by
Eqs.~(14) and (15), respectively.

Consider now initial distributions with $\dm > 0$.  For concreteness, we
examine the case where
$$
\pv0={(D-\dm)^\m \over \G(\m+1)} e^{-(D-\dm)},\qquad {\rm for\ \ }  D>\dm.
\eqnoi
$$
After substitution of this initial distribution in \back7, the
analog of \back3\ is
$$
\td u t = u^{-2\m-2} e^{-2\dm(u-1)}.
\eqnoi
$$
Solving this equation in the limit of $t \to \infty$,
the
asymptotic form  for the density is
$$
\ct = {1 \over 2\dm t} \Bigl[1-{2+2\m \over \ln(2\dm t)}+ \ldots \Bigr],
\eqnoi
$$
while the full distribution of diffusivities has the form
$$
{\pvt \over \pv0} \simeq (2\dm t)^{-{D+\dm \over 2\dm}}
\Biggl[{\ln(2\dm t) \over 2\dm}\Biggr]^{{(\m+1)D \over \dm}}.
\eqnoi
$$
Thus the concentration of the species with diffusivity $D$ decays
algebraically with a characteristic non-universal exponent
$(D+\dm)/2\dm$, but also with a logarithmic prefactor.

This distribution can be expressed in a scaling form by identifying the
appropriate scaling variable.  For this identification, it is
helpful to consider the  average diffusion coefficient which has
the time dependence $\dav - \dm \sim \dm/\ln(2\dm t)$.  This
suggests that the scaling variable is
$$
x = {\ln(2\dm t) \over 2\dm}(D-\dm).
\eqnoi
$$
With this identification and in the scaling limit
$t \to \infty$ and $D \to \dm$ but with $x$  finite, the asymptotic
distribution \back1\ can be written in the form
$$
\pvt \simeq {1 \over \G(\m+1)}(2\dm t)^{-1}{\ln(2\dm t) \over 2\dm}\Phi(x),
\eqnoi
$$
with scaling function $\Phi(x)$ again equal to $x^\mu e^{-x}$.
Thus for both $\dm = 0$ and $\dm > 0$,
the diffusivity distribution approaches the scaling form,
given by Eq.~(13) and \last, respectively, with characteristic width
$t^{-1/(2+2\m)}$ in the former case and $1/\ln(t)$ in the latter.

%%%%%%%%%%%%%%%%%%%%%%%%%%%%%%%%%%%%%%%%%%%%%%%%%%%%%%%%%%%%%%%%%%%%%%%%%%

\bigskip\centerline
{\bf III. Heterogeneous Annihilation in Low Dimensions}\medskip

In the diffusion-controlled limit, fluctuations effects govern the
long-time kinetics of heterogeneous annihilation for $d\leq 2$, in
close analogy with homogeneous single-species annihilation [1].  A
method which is ideally suited to account for the kinetics in this
regime is the Smoluchowski theory [5].  We adapt this approach to
treat the kinetics of heterogeneous annihilation when $d\leq 2$.  To
describe the method, consider first the simple and illustrative
example of a two-species system in one dimension in which a background
of identical particles with diffusivity $D$ and density $c$ contains
relatively rare impurities of diffusivity $D_I$ at concentration
$c_I\ll c$.

In the Smoluchowski approach, we first compute particle flux to a
``reference'' absorbing particle due to the rest of the particles
which comprise the uniform background.  The requisite solution to the
background concentration is $c(x,t)= c_\infty{\rm erf}(x/\sqrt{4Dt})$,
from which the particle flux at the reference particle is
$\phi=c_\infty\sqrt{D/\pi t}$.  This is identified as the effective
microscopic reaction rate, $\tilde k$.

For the two-component system of background and impurity, there are
distinct rates associated with reactions between background particles
and between the background and impurities.  In the limit of low impurity
density, we neglect the influence of background-impurity reactions on
the background density, as well as reactions among the impurities.
Under these restrictions, the rate equations become
$$
\eqalign{
\dot c &\cong - 2\tilde k_{BB} c^2 \sim -2 \sqrt{2D\over\pi t} c^2,\cr
\dot c_I & \cong - 2\tilde k_{BI} c\, c_I \sim -2 \sqrt{D+D_I\over\pi t} c\,
c_I,\cr}
\eqnoi
$$
where $\tilde k_{BB}$ and $\tilde k_{BI}$ are the effective
rates for background-background and background-impurity reactions.
Note that the numerator inside the square root involves the relative
diffusivities of the two reacting species.  From the first equation,
the background concentration vanishes as
$$
c(t)=\sqrt{\pi/32Dt}.\eqnai
$$
This deviates from the exact result [2, 3] for single-species
annihilation by the numerical factor $c(t)/c_{\rm exact}(t)=\pi/2$.
The crucial feature of the Smoluchowski approach is that
by the form of the rate equation for $c_I$, the coefficient of $c(t)$
determines the exponent of the decay of the impurity species.  We
thereby find
$$
c_I(t)\sim t^{-\sqrt{(1+\e)/8}},\eqno(26b)
$$
with $\e=D_I/D$.

As we discuss in the next section, this prediction agrees rather well
with numerical results.  Intuition for this non-universal behavior can
be gained by considering the impurity problem in the case where
particles coalesce, rather than annihilate.  This situation is
considerably simpler than that of annihilation, as the enclosing
``cage'', defined by the nearest neighbors of the impurity, evolves only
by diffusion.  Thus the many-body problem may be reduced to the
three-body problem of the impurity and its two nearest-neighbors.  By
exploiting simple geometric equivalences, this problem can be
transformed to the survival of a single random walker which diffuses
within an absorbing two-dimensional wedge whose opening angle depends on
$D/D_I$.  In this latter problem, the survival probability decays as
$t^{-\a}$, with $\a=\pi/(2\cos^{-1}(\e/(1+\e)))$.  Here $\theta=
\cos^{-1}{[\e/(1+\e)]}$ is the opening angle of the wedge [9].

Such a rigorous mapping does not exist for the impurity problem when
particles annihilate.  This is an intrinsically a many-body process,
since the cage surrounding a given particle can involve distant
neighbors.  Nevertheless, the Smoluchowski approach is essentially
identical for both the annihilation and the coalescence reactions,
except for overall factors of 2 in the rate equations.  Thus the
equivalent of Eq.~(26b) for aggregation is $c_I(t)\sim
t^{-\sqrt{(1+\e)/2}}$.  While this exponent value deviates considerably
from the exact result give above, the $\e$ dependence of the exponent
is qualitatively correct.

The mechanism underlying the non-universal decay of the impurity is the
equivalence to the survival probability of a diffusing particle inside
an absorbing interval whose length $L$ grows as $t^\b$.  Since the
probability density of the particle spreads over a spatial extent of the
order of $\sqrt{Dt}$, the survival probability decays exponentially for
$\beta<1/2$.  However, in the marginal case $\beta=1/2$, \ie, $L\sim
A\sqrt{Dt}$, the survival probability $S(t)$ decays algebraically in
time, $S(t)\sim t^{-\alpha(A)}$, but with the decay exponent dependent
on the {\it dimensionless} parameter $A$.  In the reaction process, the
size of the cage which surrounds the impurity also grows as $t^{1/2}$
for both annihilation and coalescence, since the overall density is
decaying as $t^{-1/2}$.  However, the microscopic differences in the two
reactions are important because they determine the amplitude $A$ in the
growth of the cage.  As shown above, the phenomenological Smoluchowski
treats annihilation and coalescence on the same footing and thus
provides a convenient description of the non-universal behavior.

We have also explored the case of small difference between diffusion
coefficients, $0<\d \ll 1$.  We find that intermediate asymptotic
behavior initially occurs before the final asymptotic of \last\ sets
in.  In particular, for initially equal concentrations
$$
c_{1,2} \simeq  \sqrt{\pi \over 128 D_1t}
\Bigl[1 \pm {\d \over 8}\ln(D_1t)\Bigr],
\eqnoi
$$
for $1 \ll t \ll e^{8/\d}$. The crossover time between
intermediate asymptotics \last\ and final asymptotics \back1\
diverges as $e^{8/\d}$ for $\d \to 0$.

Let us now consider heterogeneous annihilation in one dimension with a
continuous distribution of particle diffusivities.  Since the
Smoluchowski approach in one-dimension yields a flux that varies as
$\propto \sqrt{D/t}$, one finds a rate equation of the form of Eq.~(7)
with the kernel $K(D,D')\propto\sqrt{(D+D')/t}$.  The correspondence can
be made even closer by introducing the modified time variable $T =
4\sqrt{t/\pi}$ which eliminates the explicit time dependence.  The
rate equation for $\pvT$ becomes
$$
\pd \pvT T = -\pvT\indd \sqrt{D+D'}\,\pvvT.
\eqnoi
$$

We are unable to find either a scaling solution to this equation or the
exponents.  However, presumably exact values of the exponents can be
obtained by consideration of a closely related and more tractable model.
We replace the kernel $K(D,D') = \sqrt{D+D'}$ by one with the same
homogeneity degree, $K(D,D') =\sqrt{D}+\sqrt{D'}$.  The two kernels
obey the bounds ${\sqrt{D}+\sqrt{D'} \over \sqrt{2}} \le \sqrt{D+D'}\le
\sqrt{D}+\sqrt{D'}$.  This suggests that the asymptotic behavior
from the initial kernel should be identical to that predicted by
the simpler sum-root kernel $K(D,D')=\sqrt{D}+\sqrt{D'}$.

With this modified kernel, the rate equation becomes
$$
\pd \pvT T = -\pvT\, [\sqrt{D}\,\CP_0(T)+\CP_{1/2}(T)],
\eqnoi
$$
with $\CP_{1/2}(T) = \ind\, \sqrt{D}\,\pvT$.
Repeating the steps employed for the mean-field treatment,
the formal exact solution to \last\ is
$$
\pvT = \pv0 \sqrt{\cT} \exp\Biggl[-D\inTT \,\cTT\Biggr].
\eqnoi
$$
Again we substitute the scaling ansatz $P(D,t)\sim t^{\b-\a}\Phi(Dt^\b)$
into the above rate equation with the initial condition
$P(D,0)\sim D^{\mu}$ as $D\to 0$.
By expressing all factors in a scaling form
we find the two exponent relations $2\a+\b=1$ and $\a=2\b(1+\m)$ and
hence,
$$
\a = {2+2\m \over 5+4\m}, \quad \b = {1 \over 5+4\m}.
\eqnoi
$$
Here the exponents $\a$ and $\b$ refer to the time dependence of $\ct$
and $\dav$ in terms of the physical time variable $t$.

These results can be straightforwardly generalized to
arbitrary $d<2$ within the Smoluchowski approach.  A
simple calculation shows that the flux to a trap behaves as
$D^{d/2}t^{-1+d/2}$ for $d<2$.  In the rate equations, this suggests the
introduction of the modified time $T \sim (Dt)^{d/2}$.  One thereby
obtains an equation similar to \back3, but with the kernel $K(D,D') =
(D+D')^{d/2}$.  Next we replace this kernel by more tractable form,
$K(D,D') = D^{d/2}+D'^{d/2}$.  Applying a scaling analysis to the rate
equations with this kernel then gives the exponents $\a
=2d(1+\m)/[d+4(1+\m)]$ and $\b = d/[d+4(1+\m)]$.

In two dimensions, the Smoluchowski approach shows that the reaction
kernel is $K(D,D')=4\pi(D+D')/\ln(Dt)$.  In this sense, two dimensions
is the marginal case which demarcates the regime where the reaction rate
is time independent (for $d>2$) from the regime where the reaction rate
varies as a power law in time ($d<2$).  The corresponding rate equation
for $\pvt$ in $d=2$ is
$$
\pd \pvt t = -\pvt\indd {4\pi(D+D')\over \ln[(D+D')t]}\pvvt.
\eqnoi
$$
We analyze this equation in the same spirit as that employed for $d\ne
2$.  First, we approximate the slowly varying logarithmic factor by
$\ln[(D+D')t] \simeq \ln[\dav t] \simeq \a \ln(t)$ and then introduce
the auxiliary time variable $T \to 4\pi t/\a \ln(t)$, so that \last\
becomes identical in form to the mean-field rate equation, Eq.~(7).
Thus replacing $t$ by $T$, the results of the preceding section apply to
the present situation.  In particular, at asymptotically large times
$\pvt$ approaches the scaling form
$$
\pvt \simeq [t/\ln(t)]^{\b-\a} \Phi(D[t/\ln(t)]^\b),
\eqnoi
$$
with exponents and the scaling function given
by Eqs.~(14) and (15).

%%%%%%%%%%%%%%%%%%%%%%%%%%%%%%%%%%%%%%%%%%%%%%%%%%%%%%%%%%%%%%%%%%%%%%%%%%%

\bigskip\centerline{\bf IV. Numerical Results}\medskip

Consider the specific and illustrative example of a two-component
system consisting of an impurity particle with diffusivity $D_I$
within a background of identical particles with diffusivity $D$.  We
first describe the results of a time series analysis for the survival
probability of a static impurity.  Then we discuss complementary Monte
Carlo simulations for both the time dependence of the survival
probability and the eventual survival probability in a finite system.
Both techniques yield estimates for the decay exponent of the impurity
that are in good agreement with the Smoluchowski theory predictions.

The time series expansion is a general technique for evaluating
numerically the initial terms in the exact power-series expansion in
time for various observables that characterize reactive systems [10].
The technique involves the successive application of a
suitably-defined evolution operator on the initial state of the
system.  After each application of the evolution operator,
configurations for which the impurity particle has not reacted
contribute to the survival probability.  To compute the first $n$
coefficients of the time power series, it is sufficient to consider
the evolution of an $n$-site ring.  The primary limitation of the
technique is computer memory, since the number of configurations grows
exponentially with the order of the expansion.  For the case of the
static impurity, we obtained the expansion to order 21, as shown in
Table~1.  We then apply the Pad\'e analysis method suggested in
Ref.~[11] to estimate the asymptotic properties of the survival
probability.  In Figure 1, the diagonal Pad\'e approximants for the
survival probability, $S_{[n,n]}(t)$ are plotted for $n=8,9$, and 10.
These approximants are essentially identical for $t\ltwid6$ and we
conclude that they accurately describe the true survival probability
within this time range.  These Pad\'e approximants are then fitted to
the power-law form $S(t)\sim t^{-\a}$ using the fitting procedure also
suggested in [11].  We thereby estimate the exponent value $\a = 0.38
\pm 0.01$ (Table~2).

Monte Carlo simulations were also performed for the impurity problem in
one dimension.  A typical measurement involved 10 configurations of a
periodic chain of $5\times 10^5$ sites with initial concentrations
$c_I(0)=0.01$ and $c(0)=0.99$.  The data for the time dependence of both
$c_I(t)$ and $c(t)$ are quite straight on a double logarithmic plot and
exponent estimates may be made based on the local slopes of nearby data
points.  From the slopes of first-neighbor, second-neighbor, and
third-neighbor pairs, we estimate that the background species decays as
$t^{-\a_B}$, where $\a_B=0.50\pm 0.01$.  Here the error bar indicates
the magnitude of the fluctuations in the local slopes in the time regime
where the data is most linear (between $20\ltwid t\ltwid 10,000$).
Since $\a_B=1/2$, this analysis provides a useful confidence test.  For
the impurity species, the same analysis method suggests $\alpha= 0.37\pm
0.01$ (Fig.~2).

A complementary approach is based on a finite size scaling analysis for
the {\it eventual} survival probability of an impurity, $S(L)=
\lim_{t\to\infty}S(L,t)$, when starting with a single impurity and the
rest of the sites occupied by background particles on finite ring with
an odd number of sites, $L$.  We postulate that $S(L)\sim L^{-\gamma}$
(clearly $\g=1$ for $D_I=D$), while for short times, the survival
probability must be independent of the size of the system, so that
$S(L,t)\sim t^{-\a}$.  Since the two basic length scales are $\sqrt{Dt}$
and $L$, we assume that the survival probability has the scaling form
$S(L,t)\sim t^{-\a}f({\sqrt{Dt}/ L})$.  The scaling function $f$ must
have the limiting behaviors $f(x)\sim 1$ for $x\to 0$ and $f(x)\sim
x^{2\a}$ for $x\to\infty$ to account for the two asymptotic limits.
These conditions imply that $\g=2\a$.  An advantage of the measurement
of the eventual survival probability is that it typically gives more
accurate exponent estimates compared to the time-dependent observations.

We have performed Monte Carlo simulations of finite single impurity
systems until the impurity disappears or a single impurity remains to
measure the exponent $\gamma$ directly.  In Figure~3, we present the
average survival probability for $10^7$ realizations, for systems of
size $2^n+1$, with $n=1,\ldots,8$.  The data suggests the decay
exponent of $\gamma=0.75\pm0.01$, consistent with the series and
Monte Carlo estimates of $\alpha=0.38\pm0.01$.  Table 2 compares the
decay exponents found for the other values of $\epsilon=D_I/D$ to the
corresponding exponent given by the Smoluchowski approach.  The
Smoluchowski approximation yield a remarkably good description of the
asymptotic behavior of the impurity decay in the annihilation
reaction.

%%%%%%%%%%%%%%%%%%%%%%%%%%%%%%%%%%%%%%%%%%%%%%%%%%%%%%%%%%%%%%%%%%%%%%%%%%

\bigskip\centerline{\bf V. Discussion}\medskip

We have shown that there is non-universal kinetic behavior in
diffusion-controlled {\it heterogeneous} single-species annihilation.
Typically, the concentration of the species with the smallest
diffusion coefficient has the same time dependence as in {\it
homogeneous} single-species annihilation, $A+A \to 0$.  However, the
concentrations of the more mobile species decay as power laws in time,
with non-universal exponents that depend on the ratios of the
corresponding diffusivities to that of the least mobile species.
These exponents were determined by a mean-field approximation,
which should be valid for $d>2$, and by a phenomenological
Smoluchowski theory which should apply for $d<2$.

Our numerical studies focused on the limiting case where there is an
infinitesimal concentration of one species in a homogeneous background
of another species.  For this impurity problem, the behavior of the
cage which surrounds the impurity is the crucial ingredient which
determines the kinetics of the impurity.  For the coalescence reaction,
$A+A\to A$, the cage consists of the two nearest-neighbors of the
impurity.  These enclosing particles continue to undergo diffusion
even if there are reactions with other more distant particles in the
system.  This allows one to reduce the many-body system to a
three-particle system which can then be solved exactly in one
dimension.  In the annihilation reaction, a nearest-neighbor of the
impurity can disappear because of a reaction with the next-nearest
background particle.  This can cause a relatively large rearrangement
in which the cage expands to the next background particle.

The Smoluchowski approach turns out to be ideally suited for providing
a simple description of the non-universal behavior of the impurity
survival probability.  Although this approach involves an uncontrolled
approximation, the Smoluchowski method provides a simple way to
quantify the enclosing cage of the impurity for both the coalescence
and annihilation reactions.
This is a remarkable success in view of the relatively complex cage
dynamics in the annihilation reaction. From the information about
the size of the cage, the decay exponent follows directly.
In particular, for the annihilation reaction,
the exponent predictions compare well with numerical estimates.

%%%%%%%%%%%%%%%%%%%%%%%%%%%%%%%%%%%%%%%%%%%%%%%%%%%%%%%%%%%%%%%%%%%%%%%%%%%

\bigskip\centerline{\bf Acknowledgments}

We thank M.~Bramson for stimulating our interest in this problem.
We gratefully acknowledge ARO grant \#DAAH04-93-G-0021, NSF grant
\#DMR-9219845, and the Donors of The Petroleum Research
Fund, administered by the American Chemical Society,
for partial support of this research.

%%%%%%%%%%%%%%%%%%%%%%%%%%%%%%%%%%%%%%%%%%%%%%%%%%%%%%%%%%%%%%%%%%%%%%%%%

\bigskip\beginsection\centerline{References}

\refi See \eg, M.~Bramson and D.~Griffeath, \zw 53 183 1980 ;
      D. C. Torney and H. M. McConnell, \jpc 87 1941 1983 .
\refi A.~A.~Lushnikov, \jetp 64 811 1986 .
\refi Z.~R\'acz, \prl 55 1707 1985 ;
      J. G. Amar and F. Family, \pra 41 3258 1990 .
      V. Privman, \jsp 69 629 1992 .
\refi E.~Ben-Naim, S.~Redner, and F.~Leyvraz, \prl 70 1890 1993 .
\refi M.~V.~Smoluchowski, \zpc 92 215 1917 .
\refi S.~Chandrasekhar, \rmp 15 1 1943 .
\refi R.~L.~Drake, in {\it Topics in Current Aerosol Research},
      Vol. III, part 2, eds. G.~M.~Hidy and J.~R.~Brock
      (Oxford, Pergamon, 1972).
\refi E.~Ben-Naim, P.~L.~Krapivsky, and S.~Redner, \pre 50 xxx 1994 .
\refi See \eg, M.~E.~Fisher, \jsp 34 667 1984 ;
      M.~E.~Fisher and M.~P.~Gelfand, \jsp 53 175 1988 ;
      D.~ben-Avraham, \jcp 88 941 1988 ;
      M. Bramson and D. Griffeath in {\it Random walks,
      Brownian motion, and interacting particle systems:
      A festschrift in honor of Frank Spitzer},
      Eds.\ R. Durrett and H. Kesten (Birkhauser, 1991),
      and references therein; D. Considine and S. Redner, \jpa
      22 1621 1989 .
\refi S.~Song and D.~Poland, \jpa 25 3914 1992 ;
      I.~Jensen and R.~Dickman \jsp 71 89 1993 .
\refi E.~Ben-Naim and J.~Zhuo, \pre 48 2603 1993 .

%%%%%%%%%%%%%%%%%%%%%%%%%%%%%%%%%%%%%%%%%%%%%%%%%%%%%%%%%%%%%%%%%%%%%%%%%%

\vfill\eject

\bigskip\centerline{\bf Figure Captions}
\medskip
{\parindent = 0.8 true in

\item{Figure 1.}
Time dependence of the diagonal Pad\'e approximants $S_{[n,n]}(t)$
($n=8,9,10$) based on the time series for the survival probability in the
case of a static impurity.  A line of slope $-0.38$ is plotted for
reference.
\item{Figure 2.}
Monte Carlo simulation results for the survival probability of
impurity (squares) and background particles (circles) based on
10 configurations of a one-dimensional chain of 500,000 sites.
The initial impurity and background
concentrations are 0.01 and 0.99, respectively.
\item{Figure 3.}
The eventual survival probability $S(L)$ (circles) of a single static
impurity on a one-dimensional ring of $L$ sites in which the rest of the
ring is initially filled with diffusing particles.  The data shown is
based on $10^7$ realizations.  A line of slope $-0.75$ is shown for
reference.

\bigskip
\centerline{\bf Table Captions}
\medskip

\item{Table 1.}
The first 21 coefficients in the time series for the survival
probability of a static impurity, defined by $S(t)=\sum_n c_n t^n$.
\item{Table 2.}
Comparison of the numerical estimates for the exponent $\gamma$ that
characterizes the $L$ dependence of the eventual survival probability
with the corresponding predictions from the Smoluchowski theory.
Estimates are based on averaging over $10^6$ realizations.

}
\vfill\eject\bye